\documentclass[12pt]{article}
\usepackage{amsmath} 
\usepackage{amsthm} 

\newcommand{\be}{\begin{equation}}
\newcommand{\ee}{\end{equation}}
\newcommand{\bean}{\begin{eqnarray}}
\newcommand{\eean}{\end{eqnarray}}
\newcommand{\bea}{\begin{eqnarray*}}
\newcommand{\eea}{\end{eqnarray*}}
\newcommand{\bfig}{\begin{figure}}
\newcommand{\efig}{\end{figure}}
\newcommand{\ba}{\begin{array}}
\newcommand{\ea}{\end{array}}

\title{Lie point symmetries and the geodesic approximation for the Schr\"odinger-Newton equations }
\author{Oliver Robertshaw\\ and\\ Paul Tod \\
Mathematical Institute\\
University of Oxford\\
24-29 St Giles'\\
Oxford OX1 3LB\\
UK}
\begin{document}
\maketitle
\bibliographystyle{unsrt}
%
\baselineskip=24pt

\begin{abstract}
We consider two problems arising in the study of the Schr\"odinger-Newton equations. The first is to find their Lie point symmetries. The second, as an application of the first, is to investigate an approximate solution corresponding to widely separated lumps of probability. The lumps are found to move like point particles under a mutual inverse-square law of attraction. 
\end{abstract}
%

\section{Introduction}
The Schr\"odinger-Newton equations is the name given by Penrose \cite{rp2} to the coupled system of equations consisting of the Schr\"odinger equation for a wave-function $\psi$ moving in a potential $\phi$, where $\phi$ is obtained by solving the Poisson equation with source $\rho=|\psi|^2$. Specifically, the system in de-dimensionalised form, is
\begin{eqnarray}
-\nabla^2\psi+\phi\psi&=&{\rm i}\frac{\partial\psi}{\partial t}
\label{sn1}\\
\nabla^2\phi&=&|\psi|^2.
\label{sn2}
\end{eqnarray}
and it can be thought of as the Schr\"odinger equation for a particle moving in its own gravitational field.

Penrose was led to consider this system, or at least its stationary solutions, by his reflections on quantum mechanics, \cite{rp1},\cite{rp2}. The system has been independently considered several times \cite{R}, \cite{B}, sometimes under the name of the Schr\"odinger-Posson system, \cite{LTZ}. It is known to be derivable from the Einstein equations of General Relativity coupled to a complex scalar  field as a slow-motion, weak-field approximation \cite{SS}.

The system has been studied numerically \cite{B}, \cite{PMT}, \cite{HMT} and analytically \cite{MT}, \cite{tod1}. It has a conserved energy and conserves total probability. The stationary, spherically-symmetric solutions are known numerically \cite{PMT}, and the lowest-energy state has no zeroes and decays exponentially rapidly at infinity. There is evidence that `lumps' of probability can attract each other gravitationally and even orbit each other, but that only the ground-state is absolutely stable \cite{HMT}.

In this article, we consider two problems arising in the study of the Schr\"odinger-Newton equations. 

The first is to find the Lie point symmetries of the system. This is an investigation available for any system of differential equations, and a natural step in the process of understanding a system. Some symmetries are clear from the outset: the system must have the symmetries of Euclidean space, namely rotations and translations. The usual (linear) Schr\"odinger equation has a symmetry of phase transformation and a symmetry under Galilean transformation and so one expects something similar from the Schr\"odinger-Newton equations. However, as we shall see, the details of the appropriately generalised Galilean transformation contain a reflection of the Equivalence Principle from General Relativity. Finally there is a scaling symmetry.

The second problem is to find an approximation to the Schr\"odinger-Newton equations corresponding to widely separated spherically-symmetric lumps of probability. For this, we use a version of the geodesic approximation for the motion of monopoles \cite{M}. That is, we use the form of the Galilean transformation and the scaling symmetry found in the solution of the first problem to approximate the lumps with rescaled copies of the ground-state set into motion, and calculate the conserved energy as a functional of the positions, velocities and individual `probabilities' of the lumps, which play the role of masses. Then we treat this energy by the methods of classical mechanics. In this approximation, the lumps emerge as point particles moving under their mutual inverse-square attraction. This can be expected to remain a good approximation while the lumps remain widely-separated but one knows from \cite{HMT} that lumps can merge, with the radiation of probability to infinity.

The plan of the paper is as follows. In the Section 2, we review the method of finding Lie point symmetries, closely following \cite{St} and apply the method to the Schr\"odinger-Newton equations. The details of the calculation are relegated to an Appendix, but we present the resulting transformations and interpret them in this section. In Section 3, we use the symmetries from Section 2 to write down the solution corresponding to a rescaled lump in motion and integrate out the space coordinates in the energy. This leads to the classical-mechanical Lagrangian governing the motion of widely separated lumps.

\section{Lie Point Symmetries}

\subsection{General Theory}

We choose notation to follow \cite{St}, so suppose we wish to study a system of second-order partial differential equations, with independent variables $(x^i; i=1,\ldots ,M)$ and dependent variables $(u^{\beta};\beta = 1,\ldots ,N)$. We clump dependent and independent variables together as $y^a$ and label the equations of the system by an index $A$:
\[
H_A(x^i,u^\beta,u^\beta_{,n},u^\beta_{,nm}) := H_A(y^a) = 0, \label{lie:ha0}
\]
where partial derivatives are indicated by subscripts.

The {\emph{configuration space}} is the space coordinatised by $y^a$ and a {\emph{point symmetry}} is a coordinate transformation $\tilde{y}^a = \tilde{y}^a (y^b)$ of the configuration space with
$H_A(\tilde{y}^a) = 0$. A {\emph{Lie point symmetry}} is a point symmetry which arises from an infinitesimal generator, a vector field on the configuration space.

Suppose the Lie point symmetry is written in terms of its infinitesimal generator as:
\begin{eqnarray}
\tilde{x}^n &=& x^n + \varepsilon \mathbf{X} x^i + O(\varepsilon^2) =  x^n + \varepsilon \xi^n(x^i,u^\beta) + ... \cr
\tilde{u}^\alpha &=& u^\alpha + \varepsilon \mathbf{X} u^\beta + O(\varepsilon^2) = u^\alpha + \varepsilon \eta^\alpha (x^i,u^\beta) + ... \label{lie:xutransformation}
\end{eqnarray}
where
\be
\xi^n := \left.{\partial \tilde{x}^n \over \partial \varepsilon}\right|_{\varepsilon=0}, \qquad \eta^\alpha := \left.{\partial \tilde{u}^\alpha \over \partial \varepsilon}\right|_{\varepsilon=0}, \label{lie:xuepsilon}
\ee
then the infinitesimal generator is the vector field $\mathbf{X}$ on the space of $y^a$ given by
\[
\mathbf{X} = \xi^n \frac{\partial}{\partial x^n} + \eta^\alpha \frac{\partial}{\partial u^\alpha}. 
\]
For a system of second-order PDEs we need to extend the definition of ${\bf X}$ to the second jet-bundle of the configuration space:
\[
\mathbf{X} = \xi^n \frac{\partial}{\partial x^n} + \eta^\alpha \frac{\partial}{\partial u^\alpha} 
+ \eta^\alpha_{\;n} \frac{\partial}{\partial u^\alpha_{\;,n}} + \eta^\alpha_{\;nm} \frac{\partial}{\partial u^\alpha_{\;,nm}}, \label{liepoint1}
\]
where
\be
\eta^\alpha_{\;n} = \left.{\partial \tilde{u}_{,n} \over \partial \varepsilon}\right|_{\varepsilon=0}, \quad
\eta^\alpha_{\;nm} = \left.{\partial \tilde{u}_{,nm} \over \partial \varepsilon}\right|_{\varepsilon=0}.
\label{8}
\ee
We call this the prolongation of ${\bf X}$ but denote it by the same symbol.

To calculate $\eta^\alpha_{\;n}$ and $\eta^\alpha_{\;nm}$ we differentiate (\ref{lie:xutransformation}):
\begin{eqnarray}
d \tilde{u}^\alpha &=& du^\alpha + \varepsilon d\eta^\alpha + ... \cr
	&=& \left[ {\partial u^\alpha \over \partial x^i} + \varepsilon \left( {\partial \eta^\alpha \over \partial x^i} + {\partial \eta^\alpha \over \partial u^\beta}{\partial u^\beta \over \partial x^i}\right)\right] dx^i + ...  \label{9}\\ \cr
d \tilde{x}^n &=& dx^n + \varepsilon d\xi^n + ... \cr
	&=& \left[ \delta^n_m + \varepsilon \left( {\partial \xi^n \over \partial x^i} + {\partial \xi^n \over \partial u^\beta}{\partial u^\beta \over \partial x^m}\right)\right] dx^i + ...\label{10}
\end{eqnarray}
and, following \cite{St}, deduce from (\ref{8}) using (\ref{lie:xuepsilon}), (\ref{9}) and (\ref{10}) that
\begin{eqnarray}
\eta^\alpha_{\;n} &=& \eta^\alpha_{\;,n} + \eta^\alpha_{\;,\beta} u^\beta_{\;,n} - \xi^i_{\;,n} u^\alpha_{\;,i} - \xi^i_{\;,\beta} u^\beta_{\;,n} u^\alpha_{\;,i}, \label{liepoint2} \\
\eta^\alpha_{\;nm} &=& 	\eta^\alpha_{\;,nm} + \eta^\alpha_{\;,n\beta} u^\beta_{\;,m} + \eta^\alpha_{\;,m\beta} u^\beta_{\;,n} - \xi^i_{\;,nm} u^\alpha_{\;,i} \label{liepoint3} \cr
		&&	+ \; \eta^\alpha_{\;,\beta \gamma} u^\beta_{\;,n} u^\gamma_{\;,m} - u^\alpha_{\;,k} \left ( \xi^k_{\;,n\beta} u^\beta_{\;,m} + \xi^k_{\;,m\beta} u^\beta_{\;,n} \right ) \cr
		&&	- \; \xi^k_{\;,\beta\gamma} u^\beta_{\;,n} u^\gamma_{\;,m} u^\alpha_{\;,k} + \eta^\alpha_{\;,\beta} u^\beta_{\;,nm} - \xi^i_{\;,n} u^\alpha_{\;,mi} - \xi^i_{\;,m} u^\alpha_{\;,ni} \cr
		&&	- \; \xi^k_{\;,\beta} \left ( u^\alpha_{\;,k} u^\beta_{\;,nm} + u^\beta_{\;,n} u^\alpha_{\;,mk} + u^\alpha_{\;,nk} u^\beta_{\;,m} \right ). \label{liepoint3}
\end{eqnarray}
The condition that ${\bf X}$ generate a point symmetry is that the prolonged ${\bf X}$ satisfy 
\be
\mathbf{X}H_A = 0,\qquad {\rm{mod}}\; H_A.
\label{main}
\ee
The method will be to solve (\ref{main}) systematically, term by term.
\subsection{Symmetries of the time-dependent Schr\"odinger-Newton equations}

We label variables as
\begin{eqnarray*}
&x^0 = t, \quad x^1 = x, \quad x^2 = y, \quad x^3 = z,& \cr
&u^1 = u = \psi, \quad u^2 = v = \overline{\psi}, \quad u^3 = w = \phi,&
\end{eqnarray*}
and then the time-dependent Schr\"{o}dinger-Newton equations (\ref{sn1}), (\ref{sn2}) are given by
\begin{eqnarray}
H_1 &:=& i u_{,0} + u_{,11} + u_{,22} + u_{,33} - uw = 0 \label{15}\\
H_2 &:=& i v_{,0} - v_{,11} - v_{,22} - v_{,33} + vw = 0 \label{16}\\
H_3 &:=& w_{,11} + w_{,22} + w_{,33} - uv = 0.\label{17}
\end{eqnarray}

\noindent Equation (\ref{main}) gives the symmetry conditions as:
\begin{eqnarray}
\mathbf{X}H_1 &=& i\eta^u_{\;0} + \eta^u_{\;11} + \eta^u_{\;22} + \eta^u_{\;33} - \eta^u w - \eta^w u = 0 \label{snsym1}, \\
\mathbf{X}H_2 &=& i\eta^v_{\;0} - \eta^v_{\;11} - \eta^v_{\;22} - \eta^v_{\;33} + \eta^v w + \eta^w v = 0 \label{snsym2}, \\
\mathbf{X}H_3 &=& \eta^w_{\;11} + \eta^w_{\;22} + \eta^w_{\;33} - \eta^u v - \eta^v u = 0 \label{snsym3}.
\end{eqnarray}
where these are to hold modulo $H_A$.

The method is as follows: For each of the $\eta_{\;nm}^\alpha$ in the system (\ref{snsym1}-\ref{snsym3}), we compute just the terms that contain those second derivatives $u^\alpha_{\;,nm}$
that {\emph{do not}} occur in (\ref{15})-(\ref{17}). These must vanish separately, which gives rise to the first set of conditions. Using these conditions, we next calculate all terms in  $\eta^\alpha_{\;mn}$ containing second derivatives of $u_{\alpha}$, to find a second set of conditions. Finally, assisted by all previous conditions, we write the symmetry conditions including all terms, and obtain a final set of conditions. This is a straightforward but intricate process which we relegate to an appendix, just giving the solution here.

There are ten linearly independent point symmetries, including two infinite-parameter families. We may list the generators as follows:
\begin{eqnarray}
\mathbf{X_1} &=& 2t {\partial \over \partial t} + x {\partial \over \partial x} + y {\partial \over \partial y} + z {\partial \over \partial z} - 2u {\partial \over \partial u} - 2v {\partial \over \partial v} - 2w {\partial \over \partial w}, \cr
\cr
\mathbf{X_2} &=& y {\partial \over \partial x} - x {\partial \over \partial y}, \cr
\cr
\mathbf{X_3} &=& z {\partial \over \partial x} - x {\partial \over \partial z}, \cr
\cr
\mathbf{X_4} &=& z {\partial \over \partial y} - y {\partial \over \partial z}, \cr
\cr
\mathbf{X_5} &=& {\partial \over \partial t}, \qquad \mathbf{X_6} = {\partial \over \partial x}, \qquad \mathbf{X_7} = {\partial \over \partial y}, \qquad \mathbf{X_8} = {\partial \over \partial z}, \cr
\cr
\mathbf{X_9} &=&  i \Omega(t) \left( u {\partial \over \partial u} - v {\partial \over \partial v} \right) - \Omega'(t) {\partial \over \partial w}, \cr
\cr
\mathbf{X_{10}} &=& a_j(t) {\partial \over \partial x^j} + {i \over 2} x^j \left[ a_j'(t) \left( u {\partial \over \partial u} - v {\partial \over \partial v} \right) + i \, a_j''(t) {\partial \over \partial w} \right].
\label{snfamily}
\end{eqnarray}
Here ${\bf a}=(a_i(t))$ is an arbitrary 3-vector function of time and $\Omega$ is an arbitrary scalar function of time.

For the first eight transformations, the interpretation is clear. First, ${\bf X_1}$ is a scaling symmetry whose finite form can be written
\[
\tilde{t} = \lambda^{-2}t, \qquad \tilde{x}^j = \lambda^{-1}x^j, \qquad \tilde{u}^\alpha = \lambda u^\alpha,
\]
Under this transformation, the normalisation integral $I=\int|\psi|^2{\rm d}^3x$ changes: $\tilde{I}=\lambda I$. We exploit this transformation in the next section to arrive at (\ref{Psi1}) and (\ref{Phi1}).

Next, ${\bf X_2},\ldots,{\bf X_4}$ are space rotations, and ${\bf X_5},\ldots,{\bf X_8}$ are translations, in space and in time.

Next, ${\bf X_{9}}$ corresponds to a time-dependent phase-change to $\psi$ , with phase-factor $\Omega(t)$, accompanied by the subtraction of ${\Omega}'$ from $\phi$.

Finally, ${\bf X_{10}}$ is a time-dependent translation in position, accompanied by a space and time dependent phase-change in $\psi$ and an additive change to $\phi$. The infinitesimal form translates as
\[
\delta{\bf r}={\bf a},\qquad \delta\psi=i\psi(\frac{1}{2}{\bf v}.{\bf r}),\qquad \delta\phi=-\frac{1}{2}{\bf A}.{\bf r}.
\]
where ${\bf v}=\frac{d{\bf a}}{dt}$ and ${\bf A}=\frac{d{\bf v}}{dt}$. The finite form is readily found and we shall need it to derive (\ref{Psi1}) and (\ref{Phi1}).
If ${\bf A}$ is zero, this is the transformation to a frame with constant velocity, in other words a Galilean transformation. The usual Schr\"odinger equation is known to be invariant under Galilean transformation but this calculation shows that that property extends to the nonlinear Schr\"odinger-Newton system. 

If ${\bf A}\neq 0$ then we still have an invariance provided we transform the gravitational potential. This can be interpreted as a relic of the Equivalence Principle: transforming to an accelerating frame is equivalent to adding a constant gravitational field. The  Schr\"odinger-Newton equations thus satisfy the Equivalence Principle and we can think of this as being due to the fact that they can be obtained by a slow-motion, weak-field approximation from the Einstein equations of General Relativity coupled to a complex scalar field, \cite{SS}.

In the next section, we shall use ${\bf X_1}$ and ${\bf X_{10}}$ to put the stationary ground-state of total probability one into motion and reduce its total probability.

\section{The Geodesic Approximation}
We consider an approximate solution of the  Schr\"odinger-Newton equations corresponding to a number of widely separated `lumps' of probability, where each lump is described approximately by a suitably rescaled copy of the lowest energy spherically-symmetric state. (It isn't known, but is rather likely, that the lowest energy spherically-symmetric state is the actual ground state.) The lumps are rescaled copies so that the total probability is still one. 

From the previous section, we know that each lump, in the absence of all the others, can be put into uniform motion by a Galilean transformation. As a stationary solution of the  Schr\"odinger-Newton equations, it is then characterised by its total probability, position and velocity. Suppose we have initial data with $n$ lumps with corresponding parameters $(m_i, {\bf a}_i, {\bf v}_i)$ where ${\bf v}_i=\dot{\bf a}_i$ and the overdot stands for $\partial/\partial t$. The idea of the geodesic approximation is that, provided the lumps are widely separated, an approximate solution of the  Schr\"odinger-Newton equations can be found by assuming that for each lump only these parameters change. To find the equations of motion for the lumps, one substitutes the ansatz of several lumps into the energy for the theory and integrates out the spatial coordinates, to leave a classical mechanical Hamiltonian as a function of  $(m_i, {\bf a}_i, {\bf v}_i)$. This is then treated as in classical mechanics to give the equations of motion.

Suppose that $\psi_0(r){\rm e}^{-iE_0t}$ is the lowest energy spherically-symmetric state of total probability one, taken without loss of generality to be real, and $\phi_0(r)$  is the corresponding potential. Then the rescaled versions, based at ${\bf a}$ and with velocity ${\bf v}$ and total probability $m$, are given by:
\begin{eqnarray}
\Psi({\bf r},t;m,{\bf a},{\bf v})&=&m^2\psi_0(m|{\bf r}-{\bf a}|)\exp(-i(m^2E_0t-\frac{1}{2}{\bf{v.r}}+\frac{1}{4}\int|{\bf{v}}|^2{\rm d}t)
)
\label{Psi1}\\
\Phi({\bf r},t;m,{\bf a},{\bf v})&=&m^2\phi_0(m|{\bf r}-{\bf a}|).
\label{Phi1}
\end{eqnarray}
For the geodesic approximation, we assume that we have $n$ widely separated lumps so that the wave function is
\begin{eqnarray}
\Psi({\bf r},t)&=&\sum_{i=1}^n\Psi_i\nonumber\\
&=&\sum_{i=1}^n\Psi({\bf r},t;m_i,{\bf a}_i,{\bf v}_i).
\label{Psi2}
\end{eqnarray}
where, to impose the normalisation condition, we require
\[
\sum_{i=1}^nm_i=1.
\]
\label{mass}
To find the corresponding potential, we should solve the Poisson equation with source equal to $|\Psi|^2$, which is in turn a sum. However, since the ground state $\psi_0(r)$ decays exponentially rapidly at infinity and we are assuming that the lumps are widely separated, the cross-terms in this sum are everywhere small compared to the diagonal terms. Therefore we may take for $\Phi$ just the sum of the individual potentials due to the individual $\Psi_i$. Thus:
\begin{eqnarray}
\Phi({\bf r},t)&=&\sum_{i=1}^n\Phi_i\nonumber\\
&=&\sum_{i=1}^n\Phi({\bf r},t;m_i,{\bf a}_i,{\bf v}_i).
\label{Phi2}
\end{eqnarray}
The energy ${\cal{E}}$ for the  Schr\"odinger-Newton equations is as follows: 
\begin{equation}
{\cal{E}}=\int\left(\frac{1}{2}\nabla\Psi .\nabla\bar{\Psi}+\frac{1}{4}\Phi\Psi\bar{\Psi}\right){\rm d}^3x
\label{energy}
\end{equation}
with the understanding that $\Phi$ is obtained from $|\Psi|^2$ via the Poisson equation.

\noindent Now we must substitute $\Psi$ as in (\ref{Psi2}) and $\Phi$ as in (\ref{Phi2}) into (\ref{energy}) and perform the space integrations. To help with this, we need some facts about $\psi_0$, $\phi_0$ and $E_0$. We have
\begin{eqnarray}
\int|\nabla\psi_0|^2{\rm d}^3x&=&-\frac{1}{3}E_0,
\label{psi1}\\
\int\phi_0|\psi_0|^2{\rm d}^3x&=&\frac{4}{3}E_0,
\label{psi2} \\
\int|\psi_0|^2{\rm d}^3x&=&1,
\label{psi3} 
\end{eqnarray}
where the first two, given in \cite{tod1}, can be thought of as the Virial Theorem. We also need the asymptotic form of $\phi_0(r)$ which follows easily from the Poisson equation:
\begin{equation}
\phi_0(r)=-\frac{1}{4\pi|{\bf r}|}+{\rm{h.o}}.
\label{phi5}
\end{equation}
To evaluate (\ref{energy}), we split it into two terms: $S=I_1+I_2$ where
\begin{eqnarray*}
I_1&=&\int\frac{1}{2}\nabla\Psi .\nabla\bar{\Psi}{\rm d}^3x
\label{I1}\\
I_2&=&\int\frac{1}{4}\Phi\Psi\bar{\Psi}{\rm d}^3x
\label{I2}
\end{eqnarray*}
For $I_1$, we again may neglect cross terms and retain only the diagonal terms, of which a typical one is
\[\int\frac{1}{2}\nabla\Psi_1 .\nabla\bar{\Psi}_1{\rm d}^3x.\]
Substituting from (\ref{Psi1}) this is
\[\frac{1}{2}m_1^4\int|m_1\nabla\psi_0(m_1|{\bf r}-{\bf a_1}|)+\frac{1}{2}i{\bf v_1}\psi_0(m_1|{\bf r}-{\bf a_1}|)|^2{\rm d}^3r.\]
We change the variable of integration to ${\bf x}=m_1({\bf r}-{\bf a_1})$ and use (\ref{psi1}) and (\ref{psi3}) to evaluate this as
\[-\frac{1}{6}m_1^3E_0+\frac{1}{8}m_1|{\bf v_1}|^2.\]
The integral $I_2$ is composed of terms like $\frac{1}{4}\int\Phi_i|\Psi_j|^2{\rm d}x$ (neglecting terms with products like $\Psi_i\bar{\Psi_j}$ for $i\neq j$ because of the exponential fall-off). When $i=j$, we may use (\ref{psi2}) to find that this is $\frac{1}{3}m_i^3E_0$. When $i\neq j$, we use (\ref{phi5}). As long as the lumps are widely separated, we may regard the $j$-th lump as a point particle in the potential of the $i$-th to calculate this term as
\[-\frac{1}{16\pi}\frac{m_im_j}{|{\bf a_i}-{\bf a_j}|}.\]
Combining the terms for $I_1$ and $I_2$ and dropping constant multiples of $E_0$ we obtain
\[
{\cal{E}}=\frac{1}{8}\sum_{i=1}^nm_i|{\bf v_i}|^2-\frac{1}{16\pi}\sum_{i\neq j}\frac{m_im_j}{|{\bf a_i}-{\bf a_j}|.}
\label{en2}
\]
Multiplying by $4$ and switching from the Hamiltonian to the Lagrangian, we find that widely separated lumps move according to the classical mechanical Lagrangian
\[
L=\frac{1}{2}\sum_{i=1}^nm_i|{\bf v_i}|^2
+\frac{1}{4\pi}\sum_{i\neq j}\frac{m_im_j}{|{\bf a_i}-{\bf a_j}|.}
\label{lag}
\]
This is the Lagrangian for point particles of masses $m_i$ at positions ${\bf a}_i$ moving under their mutual, inverse-square law, gravitational attractions: 
the widely separated lumps of probability attract each other with `gravitational constant' $\frac{1}{4\pi}$.


\section*{Appendix}

Following the strategy of \cite{St} outlined above, we first write out the parts of $\eta^{\alpha}_{\;mn}$ which appear in (\ref{snsym1})-(\ref{snsym3}). These are 
\begin{eqnarray*}
\eta^{\alpha}_{\;nn} &=& -2 \xi^i_{\;,n} u^\alpha_{\;,ni} - 2 \xi^k_{\;,\beta} u^\beta_{\;,n} u^\alpha_{\;,nk}, \cr
\eta^{\alpha}_{\;11} &=& -2 ( \xi^0_{\;,1} u^\alpha_{\;,10} + \xi^2_{\;,1} u^\alpha_{,12} + \xi^3_{\;,1} u^\alpha_{\;,13} \cr
	&&	+ \; \xi^0_{\;,u} u_{,1} u^\alpha_{,10} + \xi^2_{\;,u} u_{,1} u^\alpha_{,12} + \xi^3_{\;,u} u_{,1} u^\alpha_{,13} \cr
	&&	+ \; \xi^0_{\;,v} v_{,1} u^\alpha_{,10} + \xi^2_{\;,v} v_{,1} u^\alpha_{,12} + \xi^3_{\;,v} v_{,1} u^\alpha_{,13} \cr
	&& 	+ \; \xi^0_{\;,w} w_{,1} u^\alpha_{,10} + \xi^2_{\;,w} w_{,1} u^\alpha_{,12} + \xi^3_{\;,w} w_{,1} u^\alpha_{,13}), \\
\eta^{\alpha}_{\;22} &=& -2 ( \xi^0_{\;,2} u_{,20} + \xi^1_{\;,2} u_{,21} + \xi^3_{\;,2} u_{,23} \cr
	&&	+ \; \xi^0_{\;,u} u_{,2} u_{,20} + \xi^1_{\;,u} u_{,2} u_{,21} + \xi^3_{\;,u} u_{,2} u_{,23} \cr
	&&	+ \; \xi^0_{\;,v} v_{,2} u_{,20} + \xi^1_{\;,v} v_{,2} u_{,21} + \xi^3_{\;,v} v_{,2} u_{,23} \cr
	&& 	+ \; \xi^0_{\;,w} w_{,2} u_{,20} + \xi^1_{\;,w} w_{,2} u_{,21} + \xi^3_{\;,w} w_{,2} u_{,23}), \\
\eta^{\alpha}_{\;33} &=& -2 ( \xi^0_{\;,3} u_{,30} + \xi^1_{\;,3} u_{,31} + \xi^2_{\;,3} u_{,32} \cr
	&&	+ \; \xi^0_{\;,u} u_{,3} u_{,30} + \xi^1_{\;,u} u_{,3} u_{,31} + \xi^2_{\;,u} u_{,3} u_{,32} \cr
	&&	+ \; \xi^0_{\;,v} v_{,3} u_{,30} + \xi^1_{\;,v} v_{,3} u_{,31} + \xi^2_{\;,v} v_{,3} u_{,32} \cr
	&& 	+ \; \xi^0_{\;,w} w_{,3} u_{,30} + \xi^1_{\;,w} w_{,3} u_{,31} + \xi^2_{\;,w} w_{,3} u_{,32}).
\end{eqnarray*}
The symmetry conditions (\ref{snsym1})-(\ref{snsym3}), retaining just second order terms, are $\eta^{\alpha}_{\;11} + \eta^{\alpha}_{\;22} + \eta^{\alpha}_{\;33} + ... = 0$, which
gives the following first set of conditions on $\xi$:
\be
\xi^i_{\;,\beta} = 0 , \label{sncond1}
\ee
\be
\xi^0_{\;,j} = 0 , \label{sncond2}
\ee
\be
\xi^1_{\;,2} = - \xi^2_{\;,1} , \; \xi^1_{\;,3} = - \xi^3_{\;,1} , \; \xi^2_{\;,3} = - \xi^3_{\;,2}, \label{sncond3}
\ee
where $i = 0,1,2,3$ and $j = 1, 2, 3$ (in future, we use $i$ and $j$ as subscripts summed over just these ranges).

Now we shall return to (\ref{liepoint3}) and calculate all second order derivatives contributing to $\eta^\alpha_{\;nn}$, using the above first set of conditions. For $\alpha=u$ we find
\begin{eqnarray*}
\eta^u_{\;11} &=& \eta^u_{\;,u} u_{,11} + \eta^u_{\;,v} v_{,11} + \eta^u_{\;,w} w_{,11} - 2 \xi^j_{\;,1} u_{,1j}, \\
\eta^u_{\;22} &=& \eta^u_{\;,u} u_{,22} + \eta^u_{\;,v} v_{,22} + \eta^u_{\;,w} w_{,22} - 2 \xi^j_{\;,2} u_{,2j}, \\
\eta^u_{\;33} &=& \eta^u_{\;,u} u_{,33} + \eta^u_{\;,v} v_{,33} + \eta^u_{\;,w} w_{,33} - 2 \xi^j_{\;,3} u_{,3j}.
\end{eqnarray*}
Inserting these into (\ref{snsym1}), considering only the contribution from second order derivatives, we find
\begin{eqnarray*}
\mathbf{X} H_1 &=& \eta^{\alpha}_{\;11} + \eta^{\alpha}_{\;22} + \eta^{\alpha}_{\;33} + ... \cr
		&=&	\left ( \eta^u_{\;u} + \eta^u_{\;v} + \eta^u_{\;w}\right ) \left ( u_{,11} + u_{,22} + u_{,33} \right ) \cr
		&&	- 2 \left ( \xi^1_{\;1} u_{,11} + \xi^2_{\;,2} u_{,22} + \xi^3_{\;,3} u_{,33} \right ) + ... 
\end{eqnarray*}
This must vanish mod $H_1$, which gives us the condition
\be
\xi^1_{\;,1} = \xi^2_{\;,2} = \xi^3_{\;,3}. \label{sncond3a}
\ee
Using this and (\ref{sncond3}) we obtain:
\begin{eqnarray}
& \xi^1_{\;,11} = - \xi^1_{\;,22} = - \xi^1_{\;,33}, & \cr
& - \xi^2_{\;,11} = \xi^2_{\;,22} = - \xi^2_{\;,33}, & \cr
& - \xi^3_{\;,11} = - \xi^3_{\;,22} = \xi^3_{\;,33}. & \label{sncond3b}
\end{eqnarray}
These ((\ref{sncond3a}) and (\ref{sncond3b})) are the second set of conditions. No extra conditions arise at this stage from considering $\mathbf{X} H_2$ or $\mathbf{X} H_3$
as they are similar at second order.

Now we write out the symmetry conditions fully, with all terms, and apply the conditions so far obtained. The first is:
\begin{eqnarray}
\mathbf{X} H_1 &=& i\eta^u_{\;0} + \eta^u_{\;11} + \eta^u_{\;22} + \eta^u_{\;33} - \eta^u w - \eta^w u \cr
	&=& 	i ( \eta^u_{\;,0} + \eta^u_{\;,u} u_{,0} + \eta^u_{\;,v} v_{,0} + \eta^u_{\;,w} w_{,0} \cr
	&&	\; - \xi^0_{\;,0} u_{,0} - \xi^1_{\;,0} u_{,1} - \xi^2_{\;,0} u_{,2} - \xi^3_{\;,0} u_{,3} ) \cr
	&& 	+ \; \eta^u_{\;,11} + \eta^u_{\;,22} + \eta^u_{\;,33} - \eta^u w - \eta^w u \cr
	&&	+ \; 2 ( \eta^u_{\;,1u} u_{,1} + \eta^u_{\;,1v} v_{,1} + \eta^u_{\;,1w} w_{,1} + \eta^u_{\;,2u} u_{,2} \cr
	&&	\; + \; \eta^u_{\;,2v} v_{,2} + \eta^u_{\;,2w} w_{,2} + \eta^u_{\;,3u} u_{,3} + \eta^u_{\;,3v} v_{,3} \cr
	&&	\; + \; \eta^u_{\;,3w} w_{,3} ) - \xi^1_{,11} u_{,1} - \xi^2_{,11} u_{,2} - \xi^3_{,11} u_{,3} \cr
	&&	- \; \xi^1_{,22} u_{,1} - \xi^2_{,22} u_{,2} - \xi^3_{,22} u_{,3} - \xi^1_{,33} u_{,1} \cr
	&&	- \; \xi^2_{,33} u_{,2} - \xi^3_{,33} u_{,3} + \eta^u_{\;,uu} \left ( u_{,1}^{\;2} + u_{,2}^{\;2} + u_{,3}^{\;2} \right ) \cr
	&&	+ \; \eta^u_{\;,vv} \left ( v_{,1}^{\;2} + v_{,2}^{\;2} + v_{,3}^{\;2} \right ) + \eta^u_{\;,ww} \left ( w_{,1}^{\;2} + w_{,2}^{\;2} + w_{,3}^{\;2} \right ) \cr
	&&	+ \; \eta^u_{\;,uv} \left ( u_{,1} v_{,1} + u_{,2} v_{,2} + u_{,3} v_{,3} \right ) \cr
	&&	+ \; \eta^u_{\;,uw} \left ( u_{,1} w_{,1} + u_{,2} w_{,2} + u_{,3} w_{,3} \right ) \cr
	&&	+ \; \eta^u_{\;,vw} \left ( v_{,1} w_{,1} + v_{,2} w_{,2} + v_{,3} w_{,3} \right ) \cr
	&&	+ \; \eta^u_{\;,u} u_{,11} + \eta^u_{\;,v} v_{,11} + \eta^u_{\;,w} w_{,11} \cr
	&&	+ \; \eta^u_{\;,u} u_{,22} + \eta^u_{\;,v} v_{,22} + \eta^u_{\;,w} w_{,22} \cr
	&&	+ \; \eta^u_{\;,u} u_{,33} + \eta^u_{\;,v} v_{,33} + \eta^u_{\;,w} w_{,33} \cr
	&&	- \; 2 \xi^1_{\;,1} u_{\;,11} - 2 \xi^2_{\;,2} u_{\;,22} - 2 \xi^3_{\;,3} u_{\;,33} \cr
	&=& 0. \label{snxh1}
\end{eqnarray}
Equation (\ref{snxh1}) must be satisfied identically in the derivatives of $u^\alpha$ mod $H_A$, so some conditions can be quickly
read off. From the coefficients of $v_{,0}$ and $w_{,0}$:
\be
\eta^u_{\;,v} = \eta^u_{\;,w} = 0, \label{sncond6}
\ee
from quadratic terms in $u^{\alpha}_{,i}$:
\be
\eta^u_{,\alpha\beta} = 0, 
\label{sncond8}
\ee
from the coefficient of $u_{,j}$:
\be
- i \xi^j_{\;,0} + 2 \eta^u_{\;,ju} - \xi^j_{,11} - \xi^j_{,22} - \xi^j_{,33} = 0, 
\label{sncond7}
\ee
for $j = 1,2,3$. The remaining terms in this equation are
\begin{eqnarray*}
i u_{,0} \left ( \eta^u_{\;,u} - \xi^0_{\;,0} \right ) + u_{,11} \left (\eta^u_{\;,u} - 2 \xi^1_{\;,1} \right) \cr
+ u_{,22} \left (\eta^u_{\;,u} - 2 \xi^2_{\;,2} \right) + u_{,33} \left (\eta^u_{\;,u} - 2 \xi^3_{\;,3} \right) \cr
+ \eta^u_{\;,11} + \eta^u_{\;,22} + \eta^u_{\;,33} -\eta^u w - \eta^w u + i \eta^u_{\;,0} &=& 0. \label{snthingyargh}
\end{eqnarray*}
again, mod $H_1$. Using $H_1 = 0$, the terms with $\eta^u_{\;,u}$ can be simplified to obtain
\begin{eqnarray*}
i \xi^0_{\;,0} u_{,0} + 2 \xi^1_{\;,1} u_{,11} + 2 \xi^2_{\;,2} u_{,22} + 2 \xi^3_{\;,3} u_{,33} - \eta^u_{\;,u} u w \cr
+ \eta^u w + \eta^w u - \eta^u_{\;,11} - \eta^u_{\;,22} - \eta^u_{\;,33} - i \eta^u_{\;,0} &=& 0, \label{sncond4}
\end{eqnarray*}
mod $H_1$. Hence
\be
\xi^0_{\;,0} = 2 \xi^1_{\;,1} = 2 \xi^2_{\;,2} = 2 \xi^3_{\;,3}, \label{sncond5}
\ee
leaving
\be
(\xi^0_{\;,0} - \eta^u_{\;,u} ) uw + \eta^u w + \eta^w u - \nabla^2 \eta^u - i \eta^u_{\;,0} = 0. \label{snetaucond}
\ee
Conditions (\ref{sncond6})
and (\ref{sncond8}) give
\be
\eta^u = u \, U(t,x,y,z) + \tilde{U} (t,x,y,z). \label{snetau}
\ee

Now, by (\ref{sncond2}) $\xi^0$ is a function of t only, so by (\ref{sncond5}) the $\xi^j$ are only first order in $x^j$:
\begin{eqnarray} 
\xi^0 &=& 2e(t), \cr
 \xi^1 &=& x e'(t) + \tilde{X}(t,y,z), \cr
 \xi^2 &=& y e'(t) + \tilde{Y}(t,x,z), \cr
 \xi^3 &=& z e'(t) + \tilde{Z}(t,x,y), \label{sncond10}
\end{eqnarray}
and (\ref{sncond7}) reduces to
\be
i \xi^j_{\;,0} = 2 \eta^u_{\;,ju}. \label{sncond11}
\ee
Note that $\xi^i$ at this stage satisfies the conformal Killing vector equation:
\[
\xi^i_{\;,j} + \xi^j_{\;,i} = 2e'(t) \, \delta_{ij} \; : i,j \neq 0.
\]
To solve, use conditions (\ref{sncond3}) and (\ref{sncond3b}) in (\ref{sncond10}) and integrate to obtain:
\begin{eqnarray}
\xi^1 &=& e'(t) x + a_2(t) y + a_3(t) z + a_6 + X(t), \cr
\xi^2 &=& - a_2(t) x + e'(t) y + a_4(t) z + a_7 + Y(t), \cr
\xi^3 &=& - a_3(t) x - a_4(t) y + e'(t) z + a_8 + Z(t), \label{snxi1}
\end{eqnarray}
where the functions of time and the constants $a_k$ are as yet unknown, but real.

Now we consider the other two symmetry conditions, taking into account conditions on $\xi$ we have found above.
The same process with $\mathbf{X}H_2$ will give
\be
i \xi^j_{\;,0} = - 2 \eta^v_{\;,jv},
\ee
\be
(\eta^v_{\;,v} - \xi^0_{\;,0}) vw + \nabla^2 \eta^v - i \eta^v_{\;,0} - \eta^v w - \eta^w v = 0, \label{snetavcond}
\ee
\be
\eta^v = v \, V(t,x,y,z) + \tilde{V}(t,x,y,z). \label{snetav}
\ee
And for $\mathbf{X}H_3$, in which time derivatives don't appear,
\begin{eqnarray*}
\mathbf{X} H_3 &=& \eta^w_{\;11} + \eta^w_{\;22} + \eta^w_{\;33} - \eta^u v - \eta^v u \cr
	&=&	\eta^w_{\;,11} + \eta^w_{\;,22} + \eta^w_{\;,33} - \eta^u v - \eta^v u \cr
	&&	+ 2 ( \eta^w_{\;,1u} u_{,1} + \eta^w_{\;,1v} v_{,1} + \eta^w_{\;,1w} w_{,1} + \eta^w_{\;,2u} u_{,2} \cr
	&&	\; + \; \eta^w_{\;,2v} v_{,2} + \eta^w_{\;,2w} w_{,2} + \eta^w_{\;,3u} u_{,3} + \eta^w_{\;,3v} v_{,3} \cr
	&&	\; + \; \eta^w_{\;,3w} w_{,3} ) + \eta^w_{\;,u} u_{,11} + \eta^w_{\;,v} v_{,11} + \eta^w_{\;,w} w_{,11} \cr
	&&	+ \; \eta^w_{\;,u} u_{,22} + \eta^w_{\;,v} v_{,22} + \eta^w_{\;,w} w_{,22} + \eta^w_{\;,u} u_{,33} \cr
	&&	+ \; \eta^w_{\;,v} v_{,33} + \eta^w_{\;,w} w_{,33} - 2 \xi^1_{\;,1} w_{,11} - 2 \xi^2_{\;,2} w_{,22} \cr
	&&	- \; 2 \xi^3_{\;,3} w_{,33}
\end{eqnarray*}
which must vanish mod $H_3$. This gives
\[
\eta^w_{\;,ww} = \eta^w_{\;,u} = \eta^w_{\;,v} = 0,
\]
\[
\eta^w_{\;,jw} = 0,
\]
\be
\left ( \eta^w_{\;,w} - \xi^0_{\;,0} \right ) uv + \nabla^2 \eta^w - \eta^u v - \eta^v u = 0, \label{snetawcond}
\ee
so that
\[
\eta^w = w \, W(t) + \tilde{W}(t,x,y,z).
\]
We will now insert this, (\ref{snetau}) and (\ref{snetav}) into equations (\ref{snetaucond}), (\ref{snetavcond}) and (\ref{snetawcond})
to calculate the $\eta^\alpha$ and $\xi^i$. Starting with (\ref{snetaucond}):
\[
uw (2e'(t) + W) + u ( \tilde{W} - \nabla^2 U - i U_{,0} ) + w \tilde{U} - \nabla^2 \tilde{U} - i \tilde{U_{,0}} = 0,
\]
whence
\be
\tilde{U} = 0, \quad W = - 2e'(t), \quad \tilde{W} = \nabla^2 U + i U_{,0}. \label{snuvw1}
\ee
Next, from (\ref{snetavcond})
\[
-vw (W+2 e'(t)) + v ( \nabla^2 V - i V_{,0} - \tilde{W} ) - w \tilde{V} + \nabla^2 \tilde{V} + i \tilde{V_{,0}} = 0,
\]
whence
\be
\tilde{V} = 0, \quad \tilde{W} = \nabla^2 V - i V_{,0}. \label{snuvw2}
\ee
And finally
\[
uv (W - 2e'(t)) + \nabla^2 \tilde{W} - uv U - uv V = 0,
\]
giving
\be
U + V = - 2e'(t), \quad \nabla^2 \tilde{W} = 0. \label{snuvw3}
\ee
We now have many conditions and there are different routes possible. From (\ref{sncond11}), (\ref{snxi1}) and (\ref{snuvw3}) we obtain
\[a_2'=a_3'=a_4'=0\]
\[U=\frac{i}{4}e''(x^2+y^2+z^2)+\frac{i}{2}(xX'+yY'+zZ')-2e'+i\Omega(t)\]
where $\Omega$ is a real function. Now recall that $V=\overline{U}$. From (\ref{snuvw1}) we obtain an expression for $\tilde{W}$, while from (\ref{snuvw2}) we obtain a different one unless $e''=0$. Thus
\[e(t)=a_1t+\frac{1}{2}a_5\]
in terms of constants $a_1$ and $a_5$. We now have enough to assemble the generators as

\begin{eqnarray*}
\xi^0 &=& 2 a_1 t + a_5, \cr
\xi^1 &=& a_1 x + a_2 y + a_3 z + a_6 + X(t), \cr
\xi^2 &=& - a_2 x + a_1 y + a_4 z + a_7 + Y(t), \cr
\xi^3 &=& - a_3 x - a_4 y + a_1 z + a_8 + Z(t), \\
\cr
\eta^u &=& u \left[ {1\over 2} i  \mathbf{r.a'} - 2 a_1 + i \Omega(t)\right], \cr
\eta^v &=& - v \left[ {1\over 2} i \mathbf{r.a'} + 2 a_1 + i \Omega(t)\right], \cr
\eta^w &=& - 2 a_1 w - {1 \over 2} \, \mathbf{r.a''} - \Omega'(t). \label{symm:xietas}
\end{eqnarray*}
where ${\bf a}(t)=(X(t),Y(t),Z(t))$ is an arbitary vector function of time and $\Omega(t)$ is an arbitary scalar function of time. In terms of these the symmetry generator is
\begin{eqnarray*}
\mathbf{X} &=& \left[ 2 a_1 t + a_5 \right] {\partial \over \partial t} + \left[ a_1 x + a_2 y + a_3 z + a_6 + X(t) \right] {\partial \over \partial x} \cr
	&&	+ \left[ - a_2 x + a_1 y + a_4 z + a_7 + Y(t) \right] {\partial \over \partial y} \cr
	&&	+ \left[ - a_3 x - a_4 y + a_1 z + a_8 + Z(t) \right] {\partial \over \partial z} \cr
	&&	+ u \left[ {1\over 2} i \mathbf{r.a'} - 2 a_1 + i \Omega(t)\right] { \partial \over \partial u} \cr
	&&	- v \left[ {1\over 2} i \mathbf{r.a'} + 2 a_1 + i \Omega(t)\right] { \partial \over \partial v} \cr
	&&	+ \left[ - 2a_1 w - {1 \over 2} \, \mathbf{r.a''} - \Omega'(t) \right] { \partial \over \partial w} \label{snsolution}
\end{eqnarray*}
which leads to (\ref{snfamily}).

\end{document}